\documentclass[a4paper,fleqn,usenatbib]{mnras}

\usepackage{mathptmx}
\usepackage{hyperref}
\usepackage{natbib}
\usepackage[T1]{fontenc}
\usepackage{ae,aecompl}

\usepackage{graphicx}	
\usepackage{amsmath}	
\usepackage{amssymb}	
\usepackage{footnote}

\title[Limits on the Mass, Velocity and Orbit of PSR~J1933$-$6211]{Limits on the Mass, Velocity and Orbit of PSR~J1933$-$6211}

\author[E. Graikou et al.]{E. Graikou,$^{1}$\thanks{E-mail: egraikou@mpifr-bonn.mpg.de}
J. P. W. Verbiest,$^{2,1}$
S. Os{\l}owski,$^{2,1,4}$
D. J. Champion,$^{1}$
T. M. Tauris,$^{1,3}$
\newauthor
F. Jankowski$^{4,5}$
and M. Kramer$^{1,6}$
\\
$^{1}$Max-Planck-Institut f\"ur Radioastronomie, Auf dem H\"ugel 69, 53121 Bonn, Germany\\
$^{2}$Fakult\"at f\"ur Physik, Universit\"at Bielefeld, Postfach 100131, 33501 Bielefeld, Germany\\
$^{3}$Argelander-Institut f\"ur Astronomie, Universit\"at Bonn, Auf dem H\"ugel 71, 53121 Bonn, Germany\\
$^{4}$Centre for Astrophysics and Supercomputing, Swinburne University of Technology, P.O. Box 218, Hawthorn, Victoria 3122, Australia\\
$^{5}$ARC Centre of Excellence for All-Sky Astrophysics (CAASTRO)\\
$^{6}$Jodrell Bank Centre for Astrophysics, The University of Manchester, Manchester, M13 9PL, United Kingdom
}

\date{Accepted XXX. Received YYY; in original form ZZZ}

\begin{document}
\label{firstpage}
\pagerange{\pageref{firstpage}--\pageref{lastpage}}
\maketitle

\begin{abstract}
We present a high-precision timing analysis of PSR~J1933$-$6211, a millisecond pulsar (MSP) with a 3.5-ms spin period and a white dwarf (WD)
companion, using data from the Parkes radio telescope. Since we have accurately measured the polarization properties of this pulsar we have applied the matrix template matching approach in which 
the times of arrival are measured using full polarimetric information. We achieved a weighted root-mean-square timing residuals (rms) of the timing residuals of 1.23 $\rm \mu s$, 15.5$\%$ improvement compared to
the total intensity timing analysis. After studying the scintillation properties of this pulsar we put constraints on the inclination angle of the system. Based on these measurements 
and on $\chi^2$ mapping we put a 2-$\sigma$ upper limit on the companion mass (0.44 M$_\odot$). Since this mass limit cannot reveal the nature of the companion we further investigate the
possibility of the companion to be a He~WD. Applying the orbital period-mass relation for such WDs, we conclude that the mass of a He~WD companion would be about 0.26$\pm$0.01~M$_\odot$ which, combined with the measured 
mass function and orbital inclination limits, would lead to a light pulsar mass $\leqslant$ 1.0 M$_\odot$. This result seems unlikely based on current neutron star formation models and we therefore conclude 
that PSR~J1933$-$6211 most likely has a CO~WD companion, which allows for a solution with a more massive pulsar.
\end{abstract}




\section{Introduction}

MSPs are rapidly rotating neutron stars that have spin periods shorter than $\sim$30$\,$ms. It is thought that they obtain 
their short spin periods from angular-momentum increases during mass transfer from their companions \citep{Bisnovatyi74, Alpar82}. The majority 
of MSPs have WD companions which can be either massive CO or ONeMg~WDs or lower-mass He~WDs. CO or ONeMg~WD companions
are less common in binary MSPs (BMSPs), originating from intermediate-mass progenitor stars which transfer their envelope mass on a short time leaving a partly recycled MSP (with a spin period of a few tens of ms; \citealt{Tauris11}). In contrast, 
He~WD companions are the most common and originate in low-mass X-ray binaries (LMXBs). Their time scale of evolution 
is long enough to allow significant amounts of matter to accrete onto the pulsar and to fully recycle them to spin periods of a few milliseconds. 

\begin{figure*}
	\centering
	\includegraphics[width=\textwidth,scale=1.0]{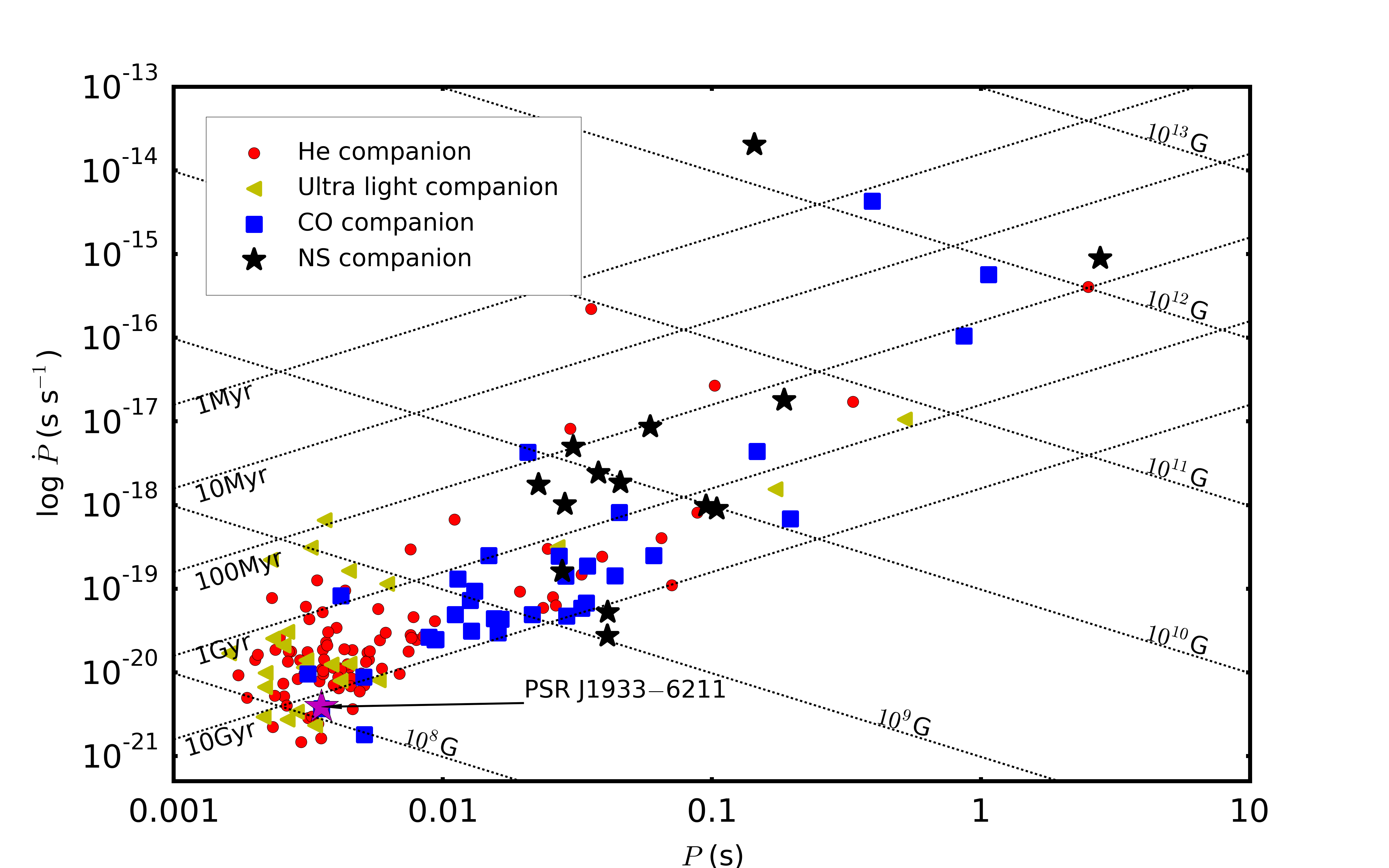}
    \caption{The position of PSR~J1933$-$6211 in the period - period derivative diagram. In the plot we present also the position of all known pulsars that have a companion.
    Circles indicate He~WD companions, squares CO~WD companions, diamonds neutron star companions and triangles ultra-light or planetary companions (data taken from the ATNF Pulsar Catalog, \citet{Manchester05}), 
    see definition  of companion stars in \citet{Tauris12}.}
    \label{fig:P_Pdot}
\end{figure*}

PSR~J1933$-$6211 was discovered in 2007 at the Parkes radio telescope in the high Galactic latitude survey \citep{Jacoby07}. Soon after the discovery it was clear that
the spin period was only 3.5 ms and the orbital period was 12.28 days, which in combination with the position of the pulsar in a spin period -- period derivative ($P$ -- $\dot{P}$)
diagram, makes this system a typical example of a BMSP (Fig.~\ref{fig:P_Pdot}). Pulsars that belong to this category, along with double neutron-star systems,
are the most well-studied so far. If BMSPs are formed in an LMXB, their resulting He~WDs follow a tight correlation between their mass and orbital period \citep{Savonije87, Tauris99}. 
This relation can be a useful tool to probe the origin of a given BMSP and place constraints on the component masses of the system. In addition, their orbital eccentricity is also predicted to be correlated with their 
orbital period \citep{Phinney92}.

Based on equation of state and evolutionary models the minimum allowed mass for rotating neutron stars is around 0.1--0.3 M$_{\odot}$ depending on their spin period \citep{Colpi93, Haensel02}. However, from current 
theories on stellar evolution and supernova (SN) explosion physics, minimum neutron star masses are expected to be larger than at least 1.0 M$_{\odot}$ \citep{tww96}. Direct pulsar mass measurements 
can be achieved with post-Keplerian parameter measurements through timing \citep{Handbook}. These measurements require an eccentric, tight system, with preferably
an edge-on orbit orientation. In edge-on orbits Shapiro-delay detection is often possible. Shapiro delay is a time-delay that arises from the photon's passage through the gravitational field of the companion star. This effect is
strongest in edge-on orbits and is typically sharply peaked, with the largest delays inflicted on observations at time of superior conjunction. Efforts to measure the Shapiro delay (and thereby the
companion and pulsar masses) therefore typically benefit from intense observations around superior conjunction, although full orbital coverage is typically needed to break covariances with other orbital parameters \citep{fw10}.

In the cases that neither the orbit is relativistic nor the inclination of the system is high, indirect limits on the mass can be derived from the mass function. The mass function defines a relation between the 
Keplerian parameters with the system inclination angle and the system masses. Unlike Keplerian parameters that can be accurately measured through timing, the inclination angle and companion mass are
very difficult to measure directly. In cases of close, massive WDs, optical observations, in combination with WD cooling models, may reveal the mass of the companion. When no direct measurements are feasible, pulsar timing 
can still be used to derive limits on the masses and orbital parameters. In those cases where the companion is a He~WD, a mass estimate can be obtained from the 
correlation between WD mass ($M_{\rm WD}$) and orbital period ($P_{\rm{b}}$).
 
Strong scintillation of the pulsar signal can be a big observational challenge. Scintillation, which has as a result that
the observed signal intensity varies, is caused by inhomogeneities and turbulence in the interstellar plasma. How strong this phenomenon is, depends on the size of 
inhomogeneities, the distance of the pulsar, the observing frequency and the peculiar velocity. PSR~J1933$-$6211, having a low dispersion measure of 11.520(2) pc cm$^{-3}$, is expected to scintillate strongly at 1.3 GHz frequency 
\citep{Rickett77}. Scintillation time scale analysis can be used in order to measure the transverse velocity of the pulsar. Scintillation transverse velocity is 
an interstellar medium effect that is caused by the pulsar's orbital velocity, the proper motion, Earth's orbital motion, the velocity of the scattering screen and the peculiar velocity. Transverse velocity measurements 
allows us to put constraints on the orbital inclination of the system \citep{Lyne1984, Ord02} leading us to better understand the masses of the system. 

In this paper we present an updated timing analysis and mass limits for PSR~J1933$-$6211. The observations and data reduction are
described in Section~\ref{sec:Observations}, while the overall timing solution and the pulsar's usefulness for pulsar timing arrays (PTAs) are discussed in
Section~\ref{sec:Timing}. Our attempts to measure the mass of
this pulsar were seriously hampered by strong scintillation around the time of superior conjunction. An analysis of these scintillation
measurements is given in Section~\ref{sec:Scintillation}. Next, we present the pulsar's proper motion and transverse velocity in Section~\ref{subsec:proper_motion}. Constraints on 
the mass of the pulsar -- and the implications for binary evolution models, are outlined in Section~\ref{sec:Mass_limits}, followed by some concluding remarks in Section~\ref{sec:Conclusions}.

\section[Observations]{Observations and data reduction}
\label{sec:Observations}

Our observations of PSR~J1933$-$6211 were performed with the Parkes 64-m radio telescope. The total span is $\sim$10 years divided into two epochs of $\sim$1 year
of observations each (Fig.~\ref{fig:residuals}). In the first epoch the pulsar is observed as part of the high Galactic latitude survey follow-up observations \citep{Jacoby07}. 
These archival data, biblically available, were recored with the Caltech Swinburne Parkes Recorder 2 (CPSR2) coherent dedispersion backend \citep{Hotan06} that provides two dual-polarization bands 
of 64 MHz, each centered on 1341 and 1405 MHz respectively. Each of the bands was split into 128 frequency channels using a polyphase filterbank.

The aim of the second epoch of observations was to precisely measure the orbital parameters of the system and the masses. The observations were performed with the
CASPSR backend \citep{Hickish16}, which provides a 256-MHz bandwidth, centered at 1382 MHz, resulting in a significant increase in sensitivity. All the observations were taken at wavelength of 21cm
with the central beam of the Parkes multibeam receiver \citep{Staveley-Smith96}.

\begin{table*}
	\centering
	\caption{The timing data characteristics, as observed with two different observing systems.}
	\label{tab:data_set}
	\begin{tabular}{l l l l l l l l l} 
		\hline
		                                                    & Data range         & Data range           & Number  & Central         & Bandwidth    & Average ToA               & EFAC     & Weigthed rms    \\
								     & (MJD)              & (Gregorian)          & of ToAs & frequency       & (MHz)        & uncertainty               &          & timing residual      \\
								     &                    &                      &         & (MHz)           &              & ($\mu$s)\footnotemark     &          & ($\mu$s)          \\
		\hline
		CPSR2						     & 52795.7--53301.4   & 2003 Jun.--2004 Oct. & 52 & 1341/1405 & 2 x 64 & 3.686 & 0.8/1.4 & 1.254 \\
		CASPSR                                              & 55676.0--56011.0   & 2011 Apr.--2012 Mar. & 96 & 1382      & 256    & 3.393 & 0.9     & 1.057 \\
		\hline
	\end{tabular}
$^1$ These values correspond to the mean ToA uncertainty for 15-min integration time observations. For more typical 1-hour integrations, the ToA uncertainty (and timing RMS) should decrease by up to a factor of two.
\end{table*}

Since one of our goals is to measure the system masses we picked our observations strategy in a way that we were sensitive to Shapiro-delay detection. This phenomenon maximally affects pulse arrival times at superior
conjunction. Superior conjunction is the point in the orbit where the pulsar is behind its companion. Consequently we defined our observing strategy to have uniform coverage through 75\% of the orbit, but
increased observing cadence during the 25\% of the orbit surrounding superior conjunction. The length of a typical observing scan was 30min.

We divided each observation into 15-min segments. Each of these 15-min sub-integrations were weighted by their signal-to-noise ratio (S/N) before the segments were integrated in time and frequency. 
The observations that provided the best S/N were combined and von Mises functions were fitted in order to create an analytic template. 
We followed the same procedure to create templates for each of the two backends for which the data were available. We measured the pulse times-of-arrival (ToAs) with two different methods. In the first method we correlated the 
analytic templates with the observations by applying goodness-of-fit statistics in the frequency domain \citep{Taylor92} and in the second method we applied the matrix template matching approach for which we used the full polarization
information of the pulsar signal in the template.


In order to optimise the sensitivity of our analysis, we performed full polarimetric calibration following the methods described by \citep{vanStraten04,vanStraten13}.
Specifically, a time-dependent receiver model was derived from long-track observations of PSR~J0437$-$4715 following the
polarimetric calibration-modelling technique described by \citet{vanStraten04}. This solution was then applied to the PSR~J1933$-$6211 data
using the measurement-equation template-matching technique of \citet{vanStraten13}. The Stokes parameters in the calibrated data follow the
conventions defined by \citet{vanStraten10}.

\begin{figure}
	\centering
	\includegraphics[width=0.75\columnwidth, angle=270]{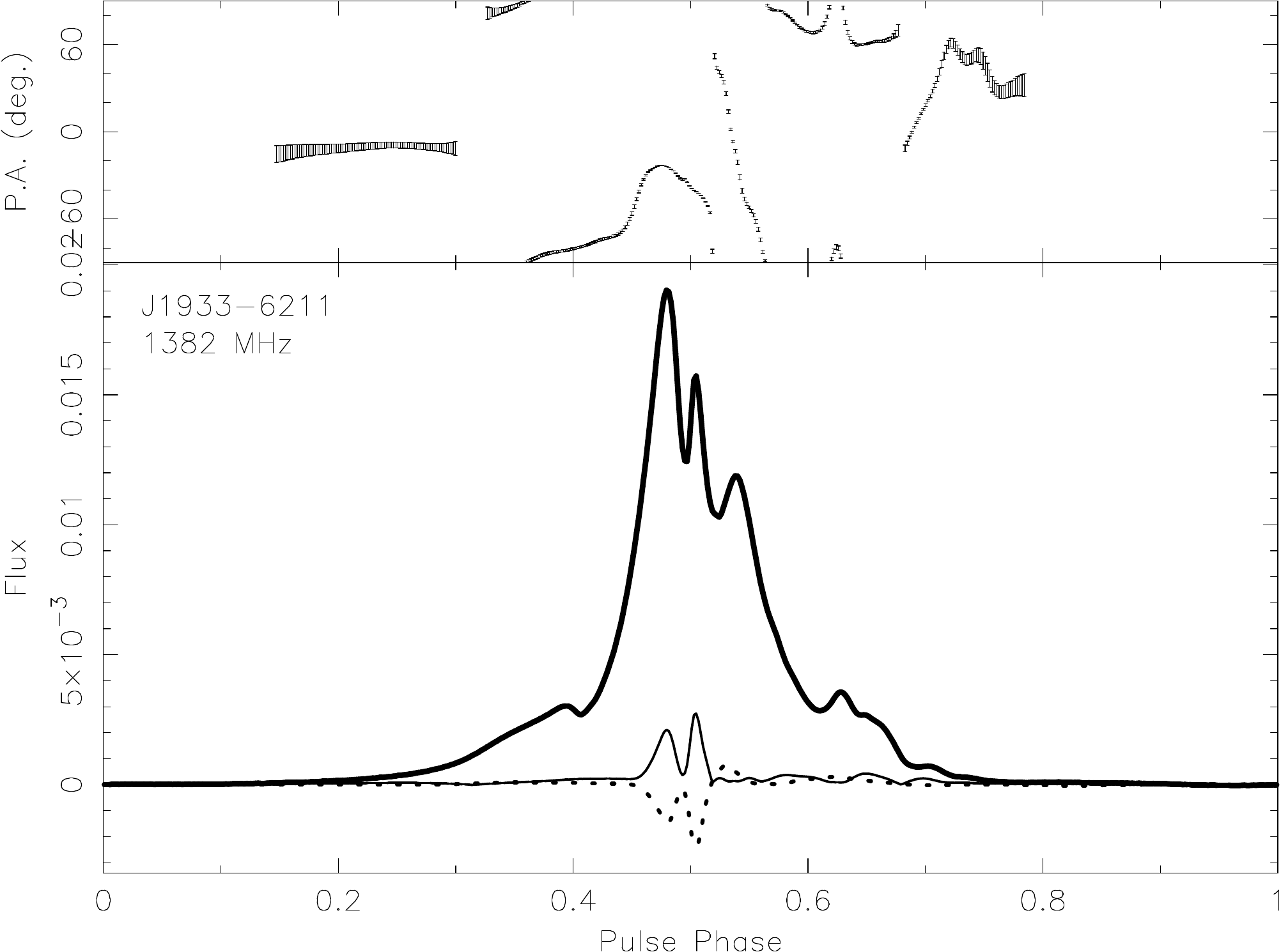}
    \caption{Average polarization calibrated pulse profile from the CASPSR backend data at 1382 MHz central frequency (averaged across a bandwidth of 256 MHz). In the plot the linear (solid) and circular polarization (dotted) profiles are presented as well as the total intensity profile (bold). 
    The flux scale is arbitrary. The fact that the main pulse is sub-divided into three individual, sharp, components, greatly aids the timing precision achievable in this system. This resolved main pulse was not 
    presented in the discovery publication \citep{Jacoby07}, which used data from the analog filterbank, which does not perform coherent dedispersion and thereby smears out the fine structure in the pulse profile.}
    \label{fig:template}
\end{figure}

\begin{figure}
	\includegraphics[width=\columnwidth,scale=1.0]{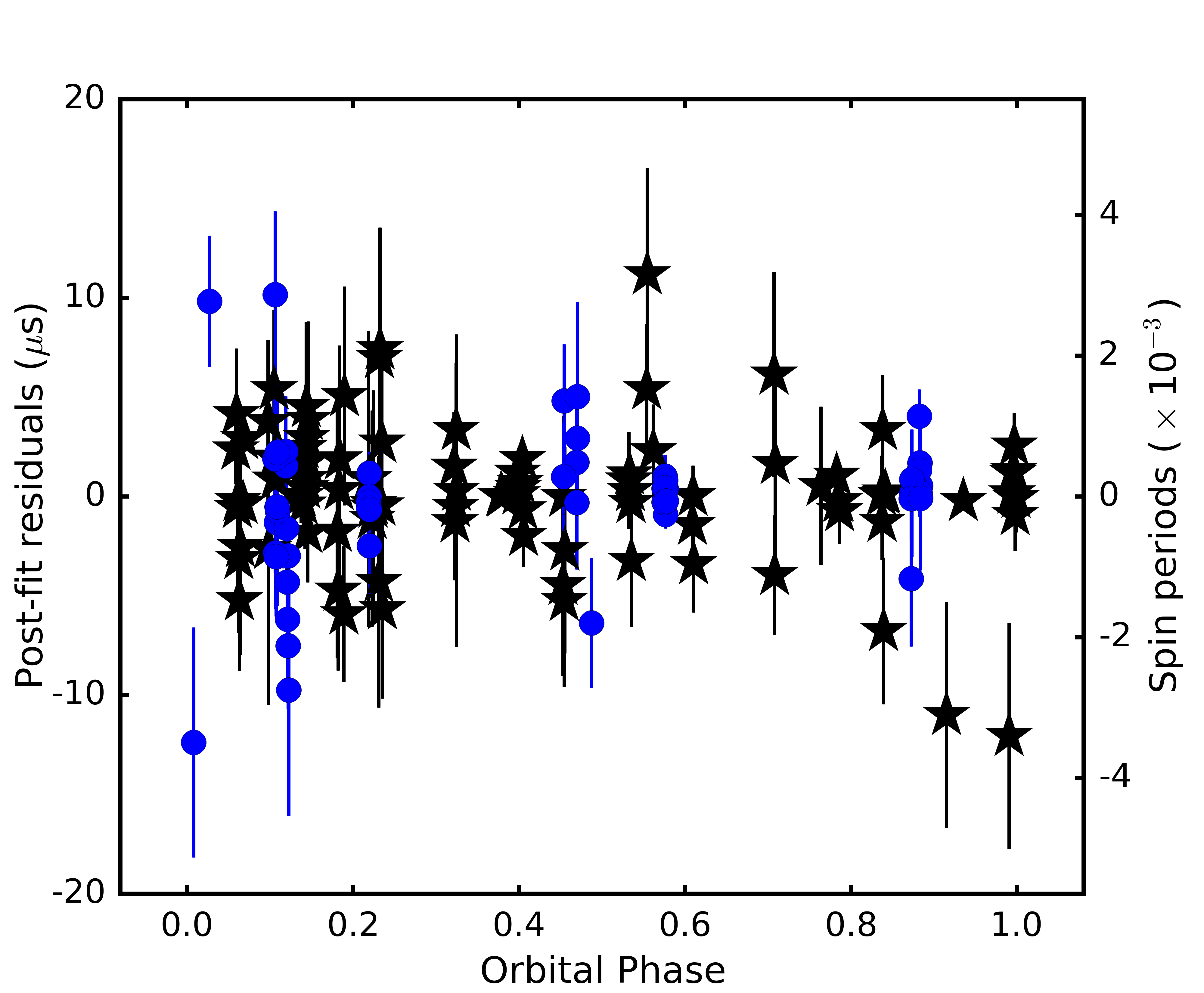}
    \caption{The timing residuals at 20 cm as a function of orbital phase, using the CPSR2 (circles) and CASPSR (stars) data. }
    \label{fig:residuals}
\end{figure}

Based on the polarized pulse shape 
(see Fig.~\ref{fig:template}), we performed the analysis of \citet{vanStraten06}, which suggests a theoretical $\sim$ 25$\%$ improvement in ToA precision, depending on the degree of polarization and the variability of the polarization
vector as a function of pulse phase, can be obtained if ToAs were derived by using the full polarization information (instead of deriving ToAs from the total-intensity profile
only, as is common in pulsar timing). To verify this result, we derived two sets of ToAs: one set based on the total-intensity pulse
profile and one based on the full polarimetric polarization, using the \citet{vanStraten06} 
timing method. In the remainder of the paper we use full polarimetric polarization residuals unless otherwise stated.

The resulting phase-connected full polarimetric polarization ToA residuals are presented in Fig.~\ref{fig:residuals}. The comparison with the total-intensity residuals led to
an overall improvement of 15.5$\%$ in timing residuals and 20\% in ToAs uncertainty.

The packages that were used for the analysis described above are \textsc{psrchive} \citep{Hotan04, vanStraten12} and \textsc{tempo2} \citep{Hobbs06}. 

\begin{table}
\centering
\caption{The timing parameters of the system that correspond to the best-fit residuals.}
\begin{tabular}{ll}
\hline
\multicolumn{2}{c}{General information} \\
\hline
MJD range\dotfill &  52795 - 53301 \\
		  &  55676 - 56011\\ 
Number of ToAs\dotfill & 148 \\
Weigthed rms timing residual ($\rm \mu s$)\dotfill &  1.23  \\ 
\hline
\multicolumn{2}{c}{Measured Quantities} \\ 
\hline
Right ascension, $\alpha$ (J2000)\dotfill & 19:33:32.42668(5) \\ 
Declination, $\delta$ (J2000)\dotfill & $-$62:11:46.876(1)\\
Proper motion in $\alpha$, $\mu_{\alpha}$ (mas\,yr$^{-1}$)\dotfill & $-$5.54(7)  \\ 
Proper motion in $\delta$, $\mu_{\delta}$ (mas\,yr$^{-1}$)\dotfill &  10.7(2) \\ 
Total proper motion (mas\,yr$^{-1}$) \dotfill & 12.1(2) \\
Spin frequency, $\nu$ (Hz)\dotfill & 282.21231361528(2) \\ 
Spin down, $\dot{\nu}$ ($\times 10^{-16}$ Hz s$^{-1}$)\dotfill & $-$3.0828(7) \\ 
Dispersion measure, $DM$ (pc cm$^{-3}$)\dotfill & 11.520(2) \\ 
Orbital period, $P_{\rm{b}}$ (d)\dotfill & 12.8194067183(8) \\
Time of ascending node, $T_{asc}$ (MJD) \dotfill &  53000.4952524(2) \\ 
Projected semi-major axis, a$_p$=a sin$i$ (lt-s)\dotfill & 12.2815745(3) \\ 
$\eta \; = \; e \; \rm sin\omega$ ($\times$ 10$^{-6}$) \dotfill & 1.36(4) \\ 
$\kappa \; = \; e \; \rm cos\omega$ ($\times$ 10$^{-6}$)\dotfill & $-$0.32(3)  \\ 
\hline
\multicolumn{2}{c}{Derived Quantities} \\
\hline
Epoch of periastron passage, $T_0$ (MJD) \dotfill &  53004.17(5) \\
Longitude of periastron, $\omega$ (deg)\dotfill & 103(1) \\
Orbital eccentricity, $e$ ($\times$ 10$^{-6}$)\dotfill & 1.40(4) \\ 
Mass function, $f(M_\odot)$ \dotfill & 0.0121034(2)  \\
\hline
\multicolumn{2}{c}{2-$\sigma$ limits from $\chi^2$ mapping} \\
\hline
Longtitude of the ascending node, $\Omega$ (deg) \dotfill & 4--148 \\
Companion mass, $M_{wd}$ (M$_{\odot}$) \dotfill & <0.44 \\
Distance, $d$ (kpc) \dotfill & >0.11 \\
\hline
\multicolumn{2}{c}{Assumptions} \\
\hline
Clock correction procedure\dotfill & TT(BIPM(2013)) \\
Solar system ephemeris model\dotfill & DE421 \\
Binary model\dotfill & ELL1 \\
\hline
\end{tabular}
Note: Figures in parentheses are  the nominal 1-$\sigma$ \textsc{tempo2} uncertainties in the least-significant digits quoted.
\label{table:ephemeris}
\end{table}

\section[Timing]{Timing solution and high-precision potential}
\label{sec:Timing}

A total of 148 ToAs were formed in order to measure the orbital, spin characteristics and the astrometry of PSR~J1933$-$6211. Since the eccentricity of the 
system is very low, we applied the ELL1 model \citep{Lange01}, for which the eccentricity, $e$, the epoch and the longitude of 
periastron passage, $T_{0}$ and $\omega$ respectively, are replaced by the Laplace-Lagrange parameters, $\eta=e\sin\omega$ and $\kappa=e\cos\omega$ and the 
time of the ascending node passage $T_{asc}$. 

We fit for pulsar parameters using least-squares fitting in order to minimize the differences between measured ToAs and the expected
arrival times derived from the analytical model \citep{Edwards06}. A phase offset was fitted between the CPSR2 and
the CASPSR data account for any difference in instrumental delays. We furthermore multiplied the ToA uncertainties with error-scaling-factors (EFACs) of 0.8 and 1.4 for the data correspond to the two CPSR2 polarization bands and 
0.9 CASPSR data respectively. These factors account for any possible underestimation of the ToA uncertainties, as described in \citet{Verbiest2016}. We note that the phase offset between the backends is only 1.71 $\mu$s. Given 
this small offset, the fact that the timing model fits both sub-sets equally well and that the fit worsens considerably when an integer number of pulse-periods get either 
added or subtracted from this offset value, clearly indicates that phase-connection was achieved across the 6-year gap.

The TT(BIPM(2013)) terrestrial time standard and the DE421 solar system ephemeris were used in order to convert from topocentric to barycenter arrival times.

The resulting timing parameters that correspond to our best-fit residuals (Fig.~\ref{fig:residuals}) are presented in Table~\ref{table:ephemeris}. 
The timing analysis led to significant measurements of the proper motion, the spin and spin-down and the keplerian parameters of the 
system. All these well measured parameters led us to further investigate the geometry and the origin of the system. 

With a weighted rms of only 1.23$\,\mu s$, PSR~J1933$-$6211 is a good candidate to be added in the Pulsar Timing Array (PTA) projects that aim to detect
gravitational waves in the nHz regime as its timing rms compares well to that of most sources in the international PTA \citep{Verbiest2016}.

\subsection[mass_constraints]{Mass constraints from timing} 
\label{subsec:4DCube}

The lack of a Shapiro delay detection and precise post-Keplerian parameter measurements, $\dot{\omega}$ and $\dot{P_{\rm{b}}}$, has as a result that with the current timing direct system-mass measurements are not possible.
Instead, we derived constraints on the masses and orientation of the system by mapping the goodness-of-fit (as determined by the $\chi^2$ value of the fit) over a four-dimensional grid covering the companion mass $M_{\rm c}$, the
inclination angle of the system $i$, the longitude of the ascending node $\Omega$ and the distance to the system $D$. At each pixel of this four-dimensional grid, these four parameters were held fixed, as were
the post-Keplerian parameters $\dot{\omega}$ and $\dot{P}_{\rm b}$ that were derived from these. (Note while the distance does not enter in the post-Keplerian parameters, it was used to determine the parallax,
which was kept fixed, and it entered in the calculation of the Kopeikin terms that define $i$ and $\Omega$; \citet{Kopeikin95, Kopeikin96}). Our grid had a step size of 0.01 M$_\odot$ in $M_c$, 0.01$^\circ$ in $i$, 1$^\circ$ in
$\Omega$ and 20 pc in distance, so our results (see Table~\ref{table:ephemeris}) are quantised at these levels.

\begin{figure}
	\includegraphics[width=\columnwidth, scale=0.8]{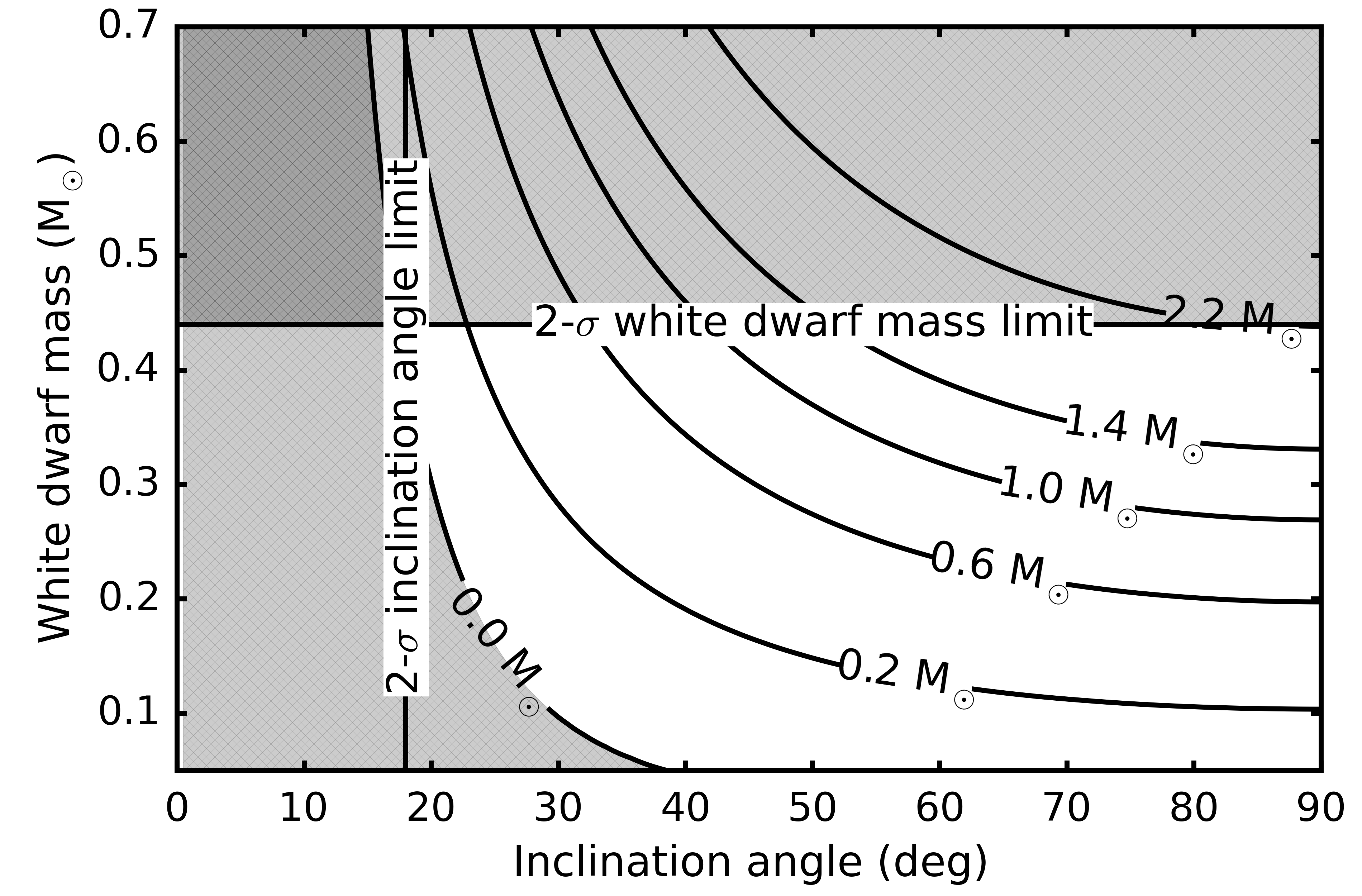}
    \caption{Mass-inclination angle diagram indicating the limits derived from timing. Specifically, the bottom-left region is excluded by the mass function and the requirement that $M_{psr} > 0$ and the 2-$\sigma$ upper
    limit on the companion mass, derived from our goodness-of-fit mapping, is indicated by the horizontal line at $M_c$ = 0.44 M$_\odot$ (see Section~\ref{subsec:4DCube}).
}
    \label{fig:white_dwarf_mass_incl}
\end{figure}

In Fig.~\ref{fig:white_dwarf_mass_incl} our results are presented, with 95$\%$ confidence limit on the companion mass. The lower limit for the WD mass (0.1 M$_\odot$)
comes directly from the mass function, evaluated for an edge-on orbit, provided that the pulsar mass is > 0.0 M$_{\odot}$. These limits do not allow us to identify the companion star as either a He or CO~WD. In Section~\ref{sec:Mass_limits} we 
revisit this question based on some alternative, more theoretical, approaches.

\subsection{Characteristic age}

The age of pulsars is key for evolutionary models. But in order to measure it precisely we should know the pulsar's spin period at birth and the braking index. In order to approximate this problem we make 
some assumptions: the initial period of a pulsar is much lower than the one that we measure today, and the pulsar loses energy due to magnetic dipole radiation with a constant B-field (braking index equal to 3). 
We can then estimate the characteristic age $\tau$:
\begin{equation}  
  \tau \; = \frac{P}{2 \; \dot{P}},
   \label{eq:characteristic_age}
\end{equation}
which is only a rough estimate of the real age (see below).

PSR~J1933$-$6211 has a very low spin period derivate, as derived from timing (see Table~\ref{table:ephemeris}). For pulsars with such low spin period derivates, effects like the
Shklovskii effect and Galactic acceleration can have a significant impact on the observed spin period derivative. In order to calculate more accurately the characteristic age we measured the contribution of 
these kinematic effects in $\dot{P}$ and we measured them. 

Both effects are quite small in PSR~J1933$-$6211, the Shklovskii effect that is calculated from:
\begin{equation}
   \left ( \frac{\dot P}{P} \right )_{\rm Shk} \; = \; 2.43 \times 10^{-21} \; \frac{\it \mu^{2}}{(\rm mass \; yr^{-1})} \; \frac{\it d}{( \rm kpc)},
   \label{eq:Shklovskii}
\end{equation}
is 1.8(3) $\times$ 10$^{-19}\,$ s s$^{-1}$. 

The Galactic acceleration contribution is defined by:
\begin{equation}
   \left (\frac{\dot{P}} {P} \right )_{\rm Gal \; acc} \; = \; \frac{\it \alpha_z \rm sin \it b}{\rm c} \rm - cos \it b \left ( \frac{\rm \Theta_{0}^{2}}{\rm cR_{0}} \right ) \left ( \rm cos \it l + \frac{\beta}{\rm sin^{2} \it l \; + \; \beta^{2}} \right ),
   \label{eq:galactic_acceleration}
\end{equation}
where the Galactic distance of the solar system: $\rm R_{0} \; = \; 8.34$ kpc, the circular rotation speed of the Galaxy at $\rm R_{0}$: $\rm \Theta_{0} \; = \; 240$ km s$^{-1}$ \citep{Reid14}, $l$ the Galactic longitude, $b$ the
Galactic latitude and $\beta \; = \; (d/\rm R_{0}) \rm cos \it b \; - \; \rm cos \it l$. It is measured to be: 9.6(6) $\times$ 10$^{-20}\,$ s s$^{-1}$. The resulting characteristic age after taken into account these effects is 
14.6(4) Gyr, which exceeds the Hubble time. Such large characteristic ages have been pointed out before \citep{Camilo94}, and are caused by the fact that the spin period derivate is too small to cause rapid changes 
in the pulsar spin period, so we observe the pulsar almost at its birth $P$ and $\dot{P}$ values \citep{Tauris12}, and thus eq.~\ref{eq:characteristic_age} above brakes down since it assumes $P_0 \; \ll \; P$, where $P_0$ is 
the birth spin period (after accretion).
Instead, for MSPs the value of $\tau$ is a measure of its remaining lifetime as a radio pulsar.

\section[Scintillation]{Scintillation measurements}
\label{sec:Scintillation}

During our observations we measured strong scintillation. The mean S/N for observations that are fully averaged in time and frequency and polarization calibrated, is ~29,
but reaches $\sim$200 during scintillation maxima, for 15-min observation scans, a factor of almost seven stronger.

We managed to observe only two complete scintles due to the length of the scintillation time scale and the fact that the individual observations were conventionally $\sim$30 minutes long. For the scintillation  
analysis we used only CASPSR data with 5-min integrations divided into 512 frequency channels (over a 256-MHz bandwidth). For each observation we removed the band edges, which corresponds to 10\% of the band. We 
manually cleaned the remaining Radio Frequency Interference (RFI) for the cases where it was needed.

\begin{figure}
	\includegraphics[width=\columnwidth,scale=1.0]{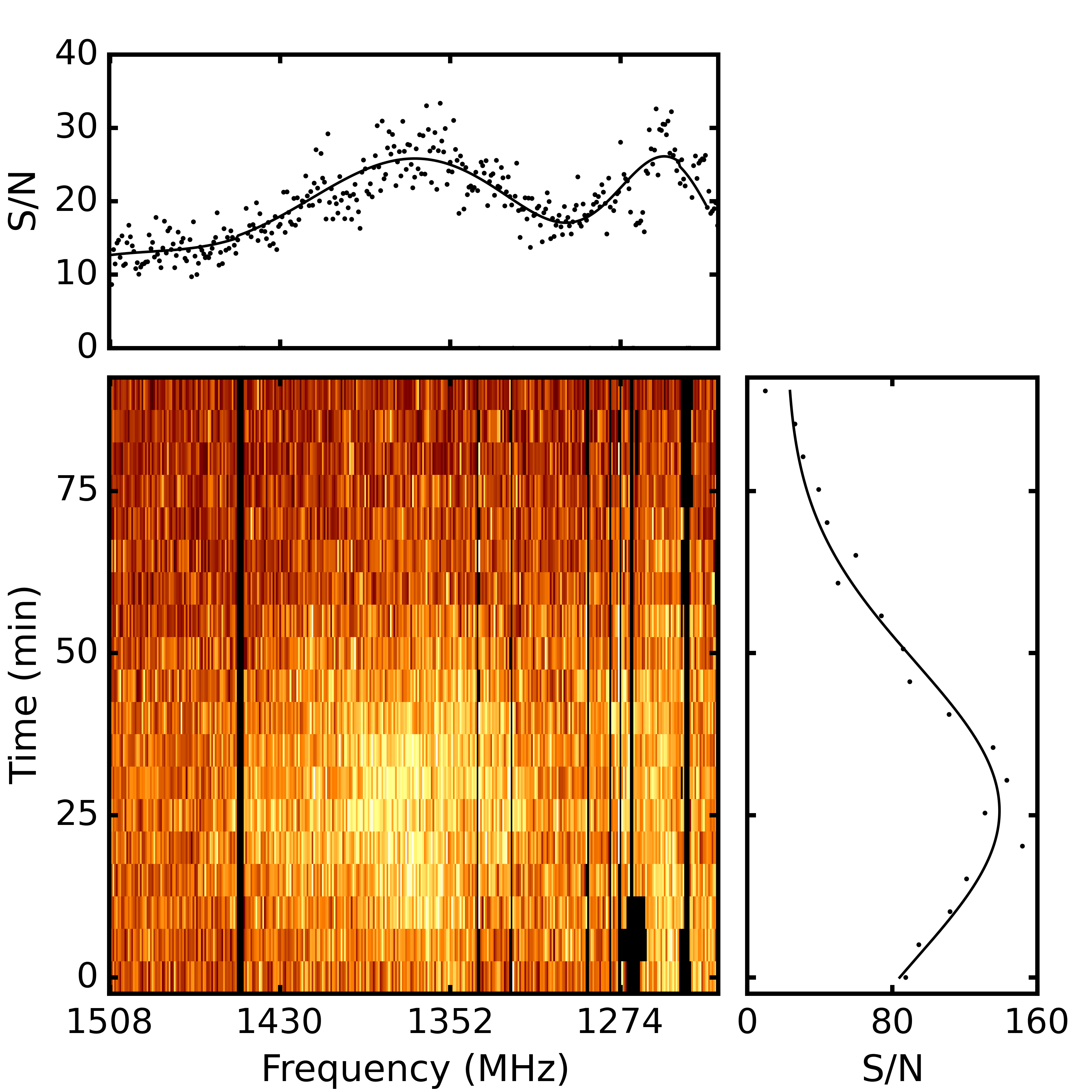}
    \caption{Dynamic spectrum of a full scintle of PSR~J1933$-$6211 as observed on 16th of May 2011. In the top and right plots, the S/N in each frequency channel and in each sub-integration, are presented respectively. The solid line
    corresponds to a Gaussian fitting line to our observation.}
    \label{fig:scintillation}
\end{figure}

In order to measure the scintillation time scale and bandwidth we fitted 1-D Gaussian functions to the frequency (respectively time) averaged intensity-series of the dynamical spectrum. The number of 
Gaussians that are used to fit the data has been decided through a Kolmogorov-Smirnov test by requiring the post-fit intensities to follow a Gaussian distribution with 95\% certainty (see Fig.~\ref{fig:scintillation}).
In Table~\ref{tab:scintillation} we present the scintillation parameters for the two available scintles as derived from the method that we explained above. 

The scintillation properties allow transverse velocity measurements. The comparison with theoretical transverse velocity and further inclination angle limits are presented below.

\begin{table}
	\centering
	\caption{Scintillation parameters and the derived scintillation speed of the two complete scintles. Both observations were taken with the CASPSR backend at 1382 MHz central frequency with 256-MHz 
	observing bandwidth divided into 512 channels and 5-min time resolution. }
	\label{tab:scintillation}
	\begin{tabular}{l | l l l} 
		\hline
		Data        		                              & 2011-05-03 & 2011-05-16\\
		\hline
		$\Delta t$ (min)                                      & 181(5)  &  108(5) \\
		$\Delta f$ (MHz)                                      & 358(8)  &  155(8)   \\
		$V_{ISS}$ based on NE2001  (km s$^{-1}$)              & 34(3)   &  38(3)  \\
		Orbital phase ($\phi$)                                & 0.2     &  0.2   \\
		\hline
	\end{tabular}
\end{table}

\subsection[proper_motion]{Transverse velocity and inclination angle limit}
\label{subsec:proper_motion}

Various studies have investigated the space velocity of pulsars. \citet{Desvignes16}, after investigating 42 MSPs pulsars, concluded that the mean transverse velocities of MSPs is 
92$\pm$10 km s$^{-1}$, a factor of 4 lower than that of young pulsars \citep{hllk05}. This difference can be understood from their dissimilar kick velocities at birth.
Such a differentiation is thought to arise from a combination of different masses of the exploding stars forming neutron stars in isolation versus in close binaries
(the latter being stripped prior to the explosion) and the absorption of the kick momentum in a binary system \citep{tb96}. 
Newborn pulsars are expected to receive a broad range of kicks -- between 0 and more than 1000 km s$^{-1}$ \citep{Janka16}. The binaries that eventually give rise to MSPs, survived the SN kick, which could mean that 
the kick often was either small or fortuitously oriented.

Pulsar space velocities can be measured through scintillation properties of the pulsar. By assuming that the diffraction screen is located in the middle between the Earth and the pulsar 
and that the scattering medium is a uniform Kolmogorov medium, we calculate the scintillation speed from \citep{Gupta94}: 
\begin{equation}
   {V_{ISS} \rm (km \; s^{-1}) \; = \; 3.85 \times 10^{4} \; \frac{\sqrt{\it d \rm (kpc) \;  \Delta \it f \rm (MHz)}}{\it f \rm (GHz) \; \Delta \it t \rm (s)}},
   \label{eq:scintillation_speed}
\end{equation}
where $d$ is the distance of the pulsar, $f$ is the observing frequency and $\Delta f$ and $\Delta t$ are the scintillation bandwidth and time scale (see Table~\ref{tab:scintillation}). 

The dispersion measure can give an estimate of the pulsar distance. We consider two such distances derived from two different models: the widely used NE2001 model \citep{CordesLazio02} and the most recent YMW16 
model \citep{Yao17}. The NE2001 model has been widely scrutinised whenever new pulsar distances were measured (see, e.g. \citealt{Matthews16, Desvignes16}), leading to the conclusion that for pulsars at 
low and medium Galactic latitudes this model gives a good distance prediction, comparable to pulsar parallax measurements. For PSR~J1933$-$6211 with Galactic latitude of $-$28.6315$^\circ$ and a dispersion 
measure equal to 11.520(2)$\,$pc cm$^{-3}$ the model-predicted distance is 0.51(7)$\,$kpc, and 0.65$\,$kpc based on NE2001 and YMW16, respectively. The scintillation speed that we measured based on NE2001 distance 
measurements is equal to 36(2) km s$^{-1}$.

The pulsar space velocity ($V_{\rm model}$), that we measured through scintillation speed, is composed of the pulsar's orbital velocity ($V_{\rm orb}$), proper motion ($V_{\rm pm}$), Earth's velocity ($V_{\rm Earth}$) and the velocity of 
the scattering screen ($V_{ISM}$).
\begin{equation}
  V_{model} \; = \; \underbrace{V_{\rm orb}}_{\alpha_p, \; P_b, \; \omega, \; i, \; e, \rm \; \phi} \; + \;  \underbrace{V_{ \rm pm}}_{\Omega, \; d, \; \mu_\alpha, \; \mu_{\delta}} \; + \; V_{\rm Earth} \; + \; V_{\rm ISM}
   \label{eq:space_velocity}
\end{equation}
The proper motion and the orbital velocity, that either added or subtracted to the proper motion, contribute the most to the pulsar space velocity, since $V_{\rm ISM}$ can be neglected \citep{Gupta95} and $V_{\rm Earth}$ is 
considered as constant throughout the observation. Since the pulsar is moving, as we observe it we track it in different positions on the sky, having as a result to 
detect differences in $V_{ISS}$. By monitoring these changes in respect to orbital phase we can obtain useful information about the geometry of the system. For example, \citet{Lyne1984} and \citet{Ord02} measured the 
inclination angles of the PSR~B0655+64 and J1141$-$6545 systems respectively, based on the orbital dependence of the scintillation speed. Unfortunately, based on our analysis, the fact that both scintles are detected at 
the same orbital phase (Table~\ref{tab:scintillation}), didn't allow us to perform a $\chi^2$ mapping in order put constraints on the inclination angle of the system. Luckily, we can further investigate the geometry 
of the system based on the timing constraints on the longitude of the ascending node ($\Omega$, see Section~\ref{subsec:4DCube} and Table~\ref{table:ephemeris}).

We compared the scintillation speed, measured with our data (Table~\ref{tab:scintillation}), with the model of the pulsar space velocity (eq.~\ref{eq:space_velocity}). The $\omega$, $P_b$, $e$, $\alpha_p$ and proper 
motion can be accurately measured through timing (Table~\ref{table:ephemeris}). Our unknowns are two: $\Omega$ and $i$. The measured orbital velocity depends only on the unknown inclination angle:
\begin{equation}
   {V_{orb} \; = \; \frac{2 \; \pi \; a_p}{\rm sin \it i \sqrt{1-e^{2}} P_{b}} \; =  \; \frac{20.8868501(5)}{\rm sin \; \it i} \; \rm km \; s^{-1}}. 
   \label{eq:orbital_speed}
\end{equation}
Higher inclination angles correspond to lower orbital velocities and as a result to lower observed scintillation speeds. The highest probable orbital velocity that can be measured for this system, in the 
0.2 orbital phase, is $\sim$120 km s$^{-1}$ and the lowest is $\sim$2 km s$^{-1}$.   

\begin{figure}
	\includegraphics[width=\columnwidth,scale=1.0]{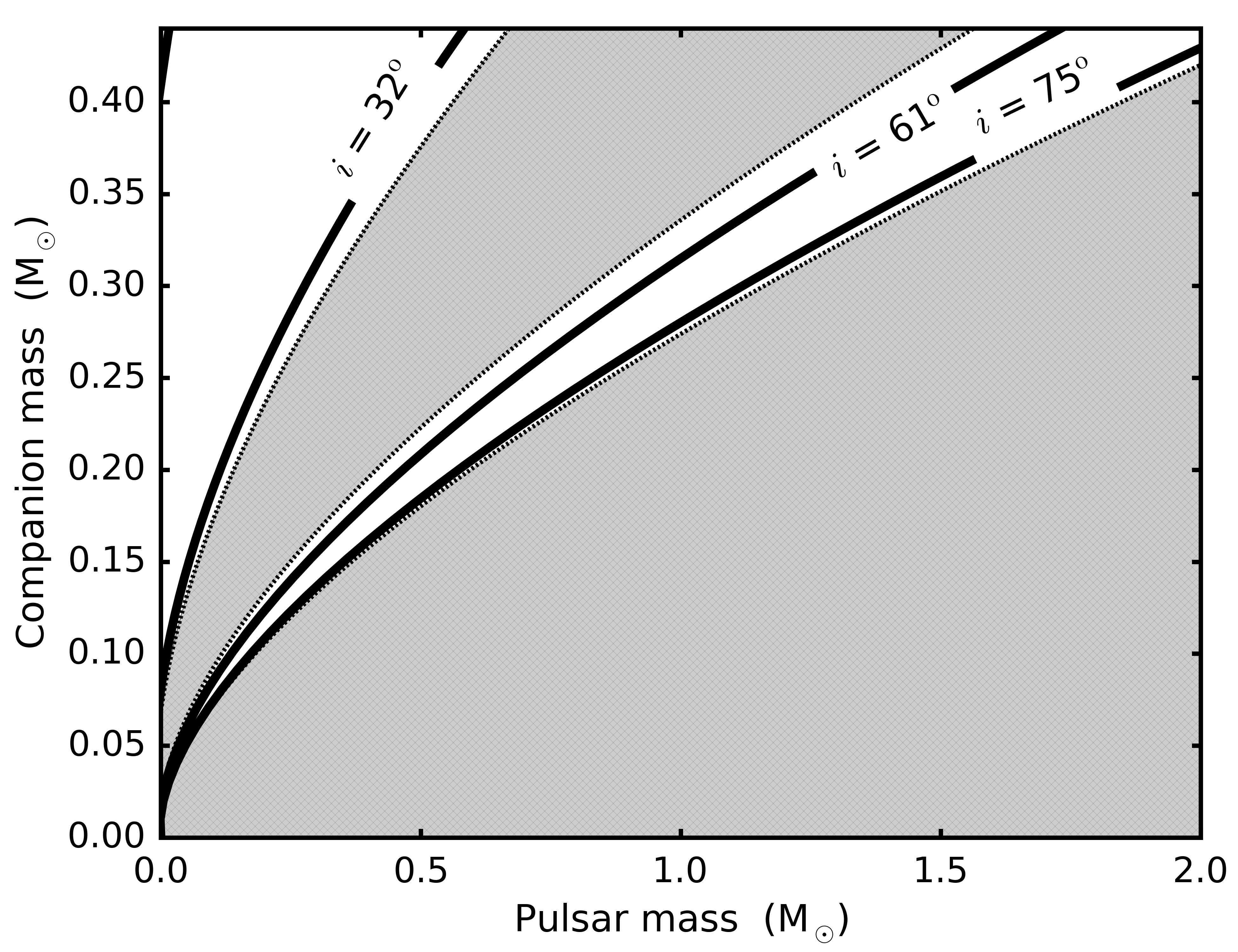}
    \caption{Pulsar mass -- companion mass diagram. The grey areas indicate inclination angles excluded with 1-$\sigma$. With the solid lines the inclination angle 1-$\sigma$ limits
    for this system are presented, as measured based on the scintillation speed. With the dotted lines the error is presented. Our two sets of solutions arise from the fact that we can not determine which of the two nodes is 
    the ascending node.}
    \label{fig:mass_mass_diagram}
\end{figure}

The other unknown is $\Omega$ for which we have a 2-$\sigma$ limit through timing (see Table~\ref{table:ephemeris}). Inside this $\Omega$ 2-$\sigma$ parameter space we find the $i$ values for which the $V_{\rm model}$ is consistent to
our scintillation speed based on our observations. Our results are presented in Fig.~\ref{fig:mass_mass_diagram}. The fact that two regions of $i$ are allowed, is due to the 180-degree ambiguity in $\Omega$, raised from the fact 
that we can not determine which of the two orbit nodes is the ascending node.

\section[Mass_limits]{Limits on the system masses}
\label{sec:Mass_limits}

In addition to the timing-based constraints presented in Section~\ref{subsec:4DCube}, for LMXBs, there are theoretical relations between the orbital parameters and the pulsar and companion masses. We will now 
consider these theoretical models and their predictions, in order to investigate the nature of the white dwarf companion.

\subsection{Orbital eccentricity}

\citet{Phinney94} found a relationship between the orbital eccentricity and orbital period for cases where the pulsar is formed after
stable mass transfer in an LMXB from a red giant filling its Roche lobe. For orbital periods of 12.8~days, the 95$\%$ lower limit on the
eccentricity as predicted from this model is significantly higher than the value that we derived from timing (Table~\ref{table:ephemeris}). This discrepancy is caused by the uncertainty in input physics applied for 
the modeling, i.e. related to tidal interactions and convection in the outer layers of the progenitor star. As a result, there is a large scatter in eccentricities for observed BMSPs with a given orbital period 
(cf. Fig. 4 in \citet{Tauris12}).

\subsection{Orbital period -- WD mass relation}
\label{subsec:Orbital_period_--_companion_mass_relation}

Different studies have investigated the relationship between the orbital period and the companion mass in 
MSP -- WD binaries which originate from LMXBs \citep{Savonije87, Rappaport95, Tauris99, Istrate14}. 
This relation has its basis in a correlation between the radius of a low-mass red-giant star and the mass of its degenerate He~core.

Numerical stellar evolution calculations have been done by \cite{Tauris99} to investigate this relationship for LMXBs with 
wide orbits (> 2 days). Based on computations of dynamically stable mass transfer in LMXBs, they obtained a relationship between the orbital period and
the WD mass for different chemical compositions of the donor star. Based on this analysis, the derived WD mass varies from 0.251 to 0.277~M$_\odot$ (Table~\ref{tab:white_dwarf_mass}).

\begin{table}
	\centering
	\caption{WD mass estimates assuming different chemical compositions of its progenitor star, based on the orbital period -- WD mass relation for He~WD companions \citep{Tauris99}
                 and given that $P_b=12.8\;{\rm d}$ for PSR~J1933$-$6211.}
	\label{tab:white_dwarf_mass}
	\begin{tabular}{l | c c} 
		\hline
		Population                     & WD mass ($\rm M_{\odot}$) \\
		\hline
		I                              & 0.251              \\
		I $\&$ II                      & 0.264              \\
		II                             & 0.277              \\
		\hline
	\end{tabular}
\end{table}

\begin{figure}
	\includegraphics[width=\columnwidth]{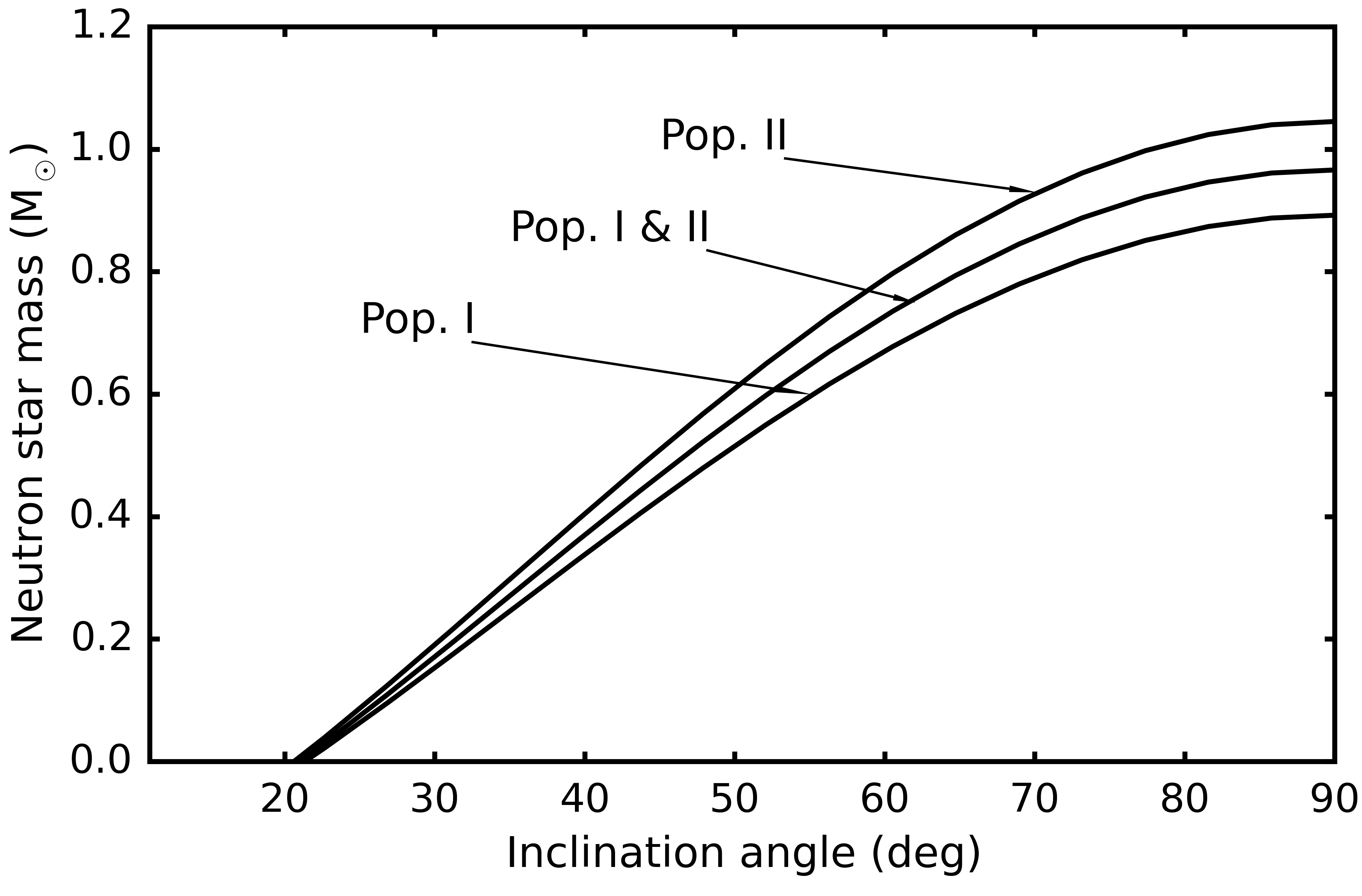}
    \caption{PSR~J1933$-$6211 mass estimate as a function of inclination angle based on the theoretical orbital period -- WD mass relations for He~WD companions (see Section~\ref{subsec:Orbital_period_--_companion_mass_relation}). 
    The three neutron star mass lines are based on different chemical compositions of the progenitor star of the WD (Table~\ref{tab:white_dwarf_mass}).}  
    \label{fig:neutron_star_mass_incl}
\end{figure}

The mass of the WD is related to the mass of the neutron star ($M_{ns}$) and the orbital inclination angle ($i$) through the mass function:
\begin{equation}
   f(M) = (M_{wd} \; \rm sin \it i)^{3} / (M_{ns}+M_{wd})^{2},
   \label{eq:mass_functiom}
\end{equation}
In Fig.~\ref{fig:neutron_star_mass_incl} we present this relation for three different WD masses, see Table~\ref{tab:white_dwarf_mass} above. The derived upper limit for the neutron star mass is 0.96 $\pm$ 0.08 M$_{\odot}$. This very 
low pulsar mass limit, can be even lower if we combine it with the inclination angle limit: $i$ < 80$^\circ$, as calculated based on the scintillation speed measurements. Neutron star masses lower than 1.1 M$_\odot$ are not consistent
with theoretical models of their formation \citep{tww96}. For this reason it is very unlikely that the WD companion would be a He~WD, and therefore we suggest that this system did not originate from an LMXB.
Hence, the WD probably has a larger mass (and quite likely a CO composition), which also implies a larger mass of the neutron star.

The formation of a fully recycled MSP, like PSR~J1933$-$6211, requires mass transfer on a long timescale which, until the discovery of PSR~J1614$-$2230
\citep{Hessels05}, has only been explained through an LMXB evolution. The discovery of the PSR~J1614$-$2230 system brought a paradox, since the pulsar is an MSP with spin period 3.2~ms and 
yet it has a CO~WD companion \citep{Demorest10}. The formation of PSR~J1614$-$2230 is explained with an intermediate-mass X-ray binary (IMXB) which evolved from 
Case~A Roche-lobe overflow (RLO) \citep{tlk11,lrp+11}. Another recently discovered candidate member of this class is PSR~J1101$-$6424 \citep{ncb+15}.
In the formation scenario of these systems the RLO begins when the progenitor
of the WD is still on the main sequence (i.e. undergoing core hydrogen burning) and therefore the timescale of the RLO is relatively long, such that 
the amount of mass that is transfered to the neutron star is sufficient to form an MSP. In case of IMXBs in a close orbit, 
the mass of the remnant CO~WD can be as low as 0.33 M$_\odot$ \citep{kw90,tvs00}. This limit is consistent with our 2-$\sigma$ WD upper limit of 0.44 M$_{\odot}$ 
derived from timing (Fig.~\ref{fig:white_dwarf_mass_incl}). Finally, we note that the orbital period of PSR~J1933$-$6211 ($12.8\;{\rm d}$) is also consistent
with an IMXB forming a CO~WD companion \citep{tvs00}. For optical constraints on the WD companion in PSR~J1933$-$6211, see \citet{Mateo17} in prep.

\section{Conclusions}
\label{sec:Conclusions}

We have presented high-precision timing of PSR~J1933$-$6211, a MSP with a WD companion, using the Parkes radio telescope. After applying the matrix template matching technique we managed to achieve ToAs
precision of 1.23~$\mu s$ and we reached an improvement of 15.5$\%$ in timing residuals, compared to the total intensity timing analysis. 
We determined precise spin and spin derivative values and the orbital parameters of the system.

The two scintles that we observed allowed us to determinate the scintillation parameters of the system. This information lead us to transverse velocity measurements of 26(2) km s$^{-1}$. Based on 
$\Omega$ 1-$\sigma$ limits, obtained from $\chi^2$ mapping,
the Keplerian parameters, measured through timing, and the transverse velocity we put limits on the inclination angle of the system. We concluded that the inclination angle is likely lower than $\sim$80$^\circ$. 

The lack of Shapiro delay, $\dot{\omega}$ and $\dot{P_{p}}$ measurements prevents direct mass measurements. For this reason, we performed $\chi^{2}$ mapping in order to put limits on the WD mass. Our 2-$\sigma$ 
upper limit for the WD mass is 0.44 $\rm M_{\odot}$. This limit corresponds to 2.2 $\rm M_{\odot}$ as an upper pulsar mass limit for a high inclination angle. Based on this limit, we cannot distinguish between the 
possibility of a CO or a He~WD. For this reason, we further investigated the possibility that the companion is a He~WD. These type of MSP systems are very well studied. 
Based on the orbital period -- WD mass relationship resulting from LMXB evolution \citep{Tauris99}, 
we derived an upper limit for the pulsar mass of 0.96~$\pm$ 0.08 M$_{\odot}$. Therefore, we conclude that if PSR~J1933$-$6211 has indeed a He~WD companion, then the mass of the pulsar is very light -- in fact, 
too light to be consistent with current stellar evolution and SN explosion physics modeling. We therefore, conclude that the companion of PSR~J1933$-$6211 is most likely a CO~WD, 
and therefore this system has an origin in an IMXB system, somewhat similar to the BMSP J1614$-$2230.

\section*{Acknowledgements}

We would like to thank Aidan Hotan and Matthew Bailes for their observing support and Willem van Straten for useful discussions. The observations 
were performed with the Parkes radio telescope: The Parkes Observatory is part of the Australian Telescope which is funded by the Commonwealth of Australia for operation as a National Facility managed by CSIRO. Parts of 
this research were conducted by the Australian Research Council Centre of Excellence for All-sky Astrophysics (CAASTRO), through project number CE110001020. 
EG acknowledges the support from IMPRS Bonn/Cologne and the Bonn--Cologne Graduate School (BCGS). SO acknowledges support from the Alexander von Humboldt Foundation and Australian Research Council 
grant Laureate Fellowship FL150100148.




\bibliographystyle{mnras}
\bibliography{./bibliography} 

\begin{thebibliography}{}
\makeatletter
\relax
\def\mn@urlcharsother{\let\do\@makeother \do\$\do\&\do\#\do\^\do\_\do\%\do\~}
\def\mn@doi{\begingroup\mn@urlcharsother \@ifnextchar [ {\mn@doi@}
  {\mn@doi@[]}}
\def\mn@doi@[#1]#2{\def\@tempa{#1}\ifx\@tempa\@empty \href
  {http://dx.doi.org/#2} {doi:#2}\else \href {http://dx.doi.org/#2} {#1}\fi
  \endgroup}
\def\mn@eprint#1#2{\mn@eprint@#1:#2::\@nil}
\def\mn@eprint@arXiv#1{\href {http://arxiv.org/abs/#1} {{\tt arXiv:#1}}}
\def\mn@eprint@dblp#1{\href {http://dblp.uni-trier.de/rec/bibtex/#1.xml}
  {dblp:#1}}
\def\mn@eprint@#1:#2:#3:#4\@nil{\def\@tempa {#1}\def\@tempb {#2}\def\@tempc
  {#3}\ifx \@tempc \@empty \let \@tempc \@tempb \let \@tempb \@tempa \fi \ifx
  \@tempb \@empty \def\@tempb {arXiv}\fi \@ifundefined
  {mn@eprint@\@tempb}{\@tempb:\@tempc}{\expandafter \expandafter \csname
  mn@eprint@\@tempb\endcsname \expandafter{\@tempc}}}

\bibitem[\protect\citeauthoryear{{Alpar}, {Cheng}, {Ruderman}  \&
  {Shaham}}{{Alpar} et~al.}{1982}]{Alpar82}
{Alpar} M.~A.,  {Cheng} A.~F.,  {Ruderman} M.~A.,   {Shaham} J.,  1982, \mn@doi
  [\nat] {10.1038/300728a0}, \href
  {http://adsabs.harvard.edu/abs/1982Natur.300..728A} {300, 728}

\bibitem[\protect\citeauthoryear{{Bisnovatyi-Kogan} \&
  {Komberg}}{{Bisnovatyi-Kogan} \& {Komberg}}{1974}]{Bisnovatyi74}
{Bisnovatyi-Kogan} G.~S.,  {Komberg} B.~V.,  1974, \sovast, \href
  {http://adsabs.harvard.edu/abs/1974SvA....18..217B} {18, 217}

\bibitem[\protect\citeauthoryear{{Camilo}, {Thorsett}  \& {Kulkarni}}{{Camilo}
  et~al.}{1994}]{Camilo94}
{Camilo} F.,  {Thorsett} S.~E.,   {Kulkarni} S.~R.,  1994, \mn@doi [\apjl]
  {10.1086/187176}, \href {http://adsabs.harvard.edu/abs/1994ApJ...421L..15C}
  {421, L15}

\bibitem[\protect\citeauthoryear{{Colpi}, {Shapiro}  \& {Teukolsky}}{{Colpi}
  et~al.}{1993}]{Colpi93}
{Colpi} M.,  {Shapiro} S.~L.,   {Teukolsky} S.~A.,  1993, \mn@doi [\apj]
  {10.1086/173118}, \href {http://adsabs.harvard.edu/abs/1993ApJ...414..717C}
  {414, 717}

\bibitem[\protect\citeauthoryear{{Cordes} \& {Lazio}}{{Cordes} \&
  {Lazio}}{2002}]{CordesLazio02}
{Cordes} J.~M.,  {Lazio} T.~J.~W.,  2002, in preprint.  (\mn@eprint {}
  {astro-ph/0207156})

\bibitem[\protect\citeauthoryear{{Demorest}, {Pennucci}, {Ransom}, {Roberts}
  \& {Hessels}}{{Demorest} et~al.}{2010}]{Demorest10}
{Demorest} P.~B.,  {Pennucci} T.,  {Ransom} S.~M.,  {Roberts} M.~S.~E.,
  {Hessels} J.~W.~T.,  2010, \mn@doi [\nat] {10.1038/nature09466}, \href
  {http://adsabs.harvard.edu/abs/2010Natur.467.1081D} {467, 1081}

\bibitem[\protect\citeauthoryear{{Desvignes} et~al.,}{{Desvignes}
  et~al.}{2016}]{Desvignes16}
{Desvignes} G.,  et~al., 2016, \mn@doi [\mnras] {10.1093/mnras/stw483}, \href
  {http://adsabs.harvard.edu/abs/2016MNRAS.458.3341D} {458, 3341}

\bibitem[\protect\citeauthoryear{{Edwards}, {Hobbs}  \& {Manchester}}{{Edwards}
  et~al.}{2006}]{Edwards06}
{Edwards} R.~T.,  {Hobbs} G.~B.,   {Manchester} R.~N.,  2006, \mn@doi [\mnras]
  {10.1111/j.1365-2966.2006.10870.x}, \href
  {http://adsabs.harvard.edu/abs/2006MNRAS.372.1549E} {372, 1549}

\bibitem[\protect\citeauthoryear{{Freire} \& {Wex}}{{Freire} \&
  {Wex}}{2010}]{fw10}
{Freire} P.~C.~C.,  {Wex} N.,  2010, \mn@doi [\mnras]
  {10.1111/j.1365-2966.2010.17319.x}, \href
  {http://adsabs.harvard.edu/abs/2010MNRAS.409..199F} {409, 199}

\bibitem[\protect\citeauthoryear{{Gupta}}{{Gupta}}{1995}]{Gupta95}
{Gupta} Y.,  1995, \mn@doi [\apj] {10.1086/176258}, \href
  {http://adsabs.harvard.edu/abs/1995ApJ...451..717G} {451, 717}

\bibitem[\protect\citeauthoryear{{Gupta}, {Rickett}  \& {Lyne}}{{Gupta}
  et~al.}{1994}]{Gupta94}
{Gupta} Y.,  {Rickett} B.~J.,   {Lyne} A.~G.,  1994, \mn@doi [\mnras]
  {10.1093/mnras/269.4.1035}, \href
  {http://adsabs.harvard.edu/abs/1994MNRAS.269.1035G} {269, 1035}

\bibitem[\protect\citeauthoryear{{Haensel}, {Zdunik}  \& {Douchin}}{{Haensel}
  et~al.}{2002}]{Haensel02}
{Haensel} P.,  {Zdunik} J.~L.,   {Douchin} F.,  2002, \mn@doi [\aap]
  {10.1051/0004-6361:20020131}, \href
  {http://adsabs.harvard.edu/abs/2002A%26A...385..301H} {385, 301}

\bibitem[\protect\citeauthoryear{{Hessels}, {Ransom}, {Roberts}, {Kaspi},
  {Livingstone}, {Tam}  \& {Crawford}}{{Hessels} et~al.}{2005}]{Hessels05}
{Hessels} J.,  {Ransom} S.,  {Roberts} M.,  {Kaspi} V.,  {Livingstone} M.,
  {Tam} C.,   {Crawford} F.,  2005, in {Rasio} F.~A.,  {Stairs} I.~H.,  eds,
  Astronomical Society of the Pacific Conference Series Vol. 328, Binary Radio
  Pulsars. p.~395 (\mn@eprint {} {astro-ph/0404167})

\bibitem[\protect\citeauthoryear{{Hickish} et~al.,}{{Hickish}
  et~al.}{2016}]{Hickish16}
{Hickish} J.,  et~al., 2016, \mn@doi [Journal of Astronomical Instrumentation]
  {10.1142/S2251171716410014}, \href
  {http://adsabs.harvard.edu/abs/2016JAI.....541001H} {5, 1641001}

\bibitem[\protect\citeauthoryear{{Hobbs}, {Lorimer}, {Lyne}  \&
  {Kramer}}{{Hobbs} et~al.}{2005}]{hllk05}
{Hobbs} G.,  {Lorimer} D.~R.,  {Lyne} A.~G.,   {Kramer} M.,  2005, \mn@doi
  [\mnras] {10.1111/j.1365-2966.2005.09087.x}, \href
  {http://adsabs.harvard.edu/abs/2005MNRAS.360..974H} {360, 974}

\bibitem[\protect\citeauthoryear{{Hobbs}, {Edwards}  \& {Manchester}}{{Hobbs}
  et~al.}{2006}]{Hobbs06}
{Hobbs} G.~B.,  {Edwards} R.~T.,   {Manchester} R.~N.,  2006, \mn@doi [\mnras]
  {10.1111/j.1365-2966.2006.10302.x}, \href
  {http://adsabs.harvard.edu/abs/2006MNRAS.369..655H} {369, 655}

\bibitem[\protect\citeauthoryear{{Hotan}, {van Straten}  \&
  {Manchester}}{{Hotan} et~al.}{2004}]{Hotan04}
{Hotan} A.~W.,  {van Straten} W.,   {Manchester} R.~N.,  2004, \mn@doi [\pasa]
  {10.1071/AS04022}, \href {http://adsabs.harvard.edu/abs/2004PASA...21..302H}
  {21, 302}

\bibitem[\protect\citeauthoryear{{Hotan}, {Bailes}  \& {Ord}}{{Hotan}
  et~al.}{2006}]{Hotan06}
{Hotan} A.~W.,  {Bailes} M.,   {Ord} S.~M.,  2006, \mn@doi [\mnras]
  {10.1111/j.1365-2966.2006.10394.x}, \href
  {http://adsabs.harvard.edu/abs/2006MNRAS.369.1502H} {369, 1502}

\bibitem[\protect\citeauthoryear{{Istrate}, {Tauris}  \& {Langer}}{{Istrate}
  et~al.}{2014}]{Istrate14}
{Istrate} A.~G.,  {Tauris} T.~M.,   {Langer} N.,  2014, \mn@doi [\aap]
  {10.1051/0004-6361/201424680}, \href
  {http://adsabs.harvard.edu/abs/2014A%26A...571A..45I} {571, A45}

\bibitem[\protect\citeauthoryear{{Jacoby}, {Bailes}, {Ord}, {Knight}  \&
  {Hotan}}{{Jacoby} et~al.}{2007}]{Jacoby07}
{Jacoby} B.~A.,  {Bailes} M.,  {Ord} S.~M.,  {Knight} H.~S.,   {Hotan} A.~W.,
  2007, \mn@doi [\apj] {10.1086/509312}, \href
  {http://adsabs.harvard.edu/abs/2007ApJ...656..408J} {656, 408}

\bibitem[\protect\citeauthoryear{{Janka}}{{Janka}}{2016}]{Janka16}
{Janka} H.-T.,  2016, preprint, \href
  {http://adsabs.harvard.edu/abs/2016arXiv161107562J} {} (\mn@eprint {arXiv}
  {1611.07562})

\bibitem[\protect\citeauthoryear{{Kippenhahn} \& {Weigert}}{{Kippenhahn} \&
  {Weigert}}{1990}]{kw90}
{Kippenhahn} R.,  {Weigert} A.,  1990, {Stellar Structure and Evolution}.
Springer-Verlag Berlin Heidelberg New York.

\bibitem[\protect\citeauthoryear{{Kopeikin}}{{Kopeikin}}{1995}]{Kopeikin95}
{Kopeikin} S.~M.,  1995, \mn@doi [\apjl] {10.1086/187731}, \href
  {http://adsabs.harvard.edu/abs/1995ApJ...439L...5K} {439, L5}

\bibitem[\protect\citeauthoryear{{Kopeikin}}{{Kopeikin}}{1996}]{Kopeikin96}
{Kopeikin} S.~M.,  1996, \mn@doi [\apjl] {10.1086/310201}, \href
  {http://adsabs.harvard.edu/abs/1996ApJ...467L..93K} {467, L93}

\bibitem[\protect\citeauthoryear{{Lange}, {Camilo}, {Wex}, {Kramer}, {Backer},
  {Lyne}  \& {Doroshenko}}{{Lange} et~al.}{2001}]{Lange01}
{Lange} C.,  {Camilo} F.,  {Wex} N.,  {Kramer} M.,  {Backer} D.~C.,  {Lyne}
  A.~G.,   {Doroshenko} O.,  2001, \mn@doi [\mnras]
  {10.1046/j.1365-8711.2001.04606.x}, \href
  {http://adsabs.harvard.edu/abs/2001MNRAS.326..274L} {326, 274}

\bibitem[\protect\citeauthoryear{{Lin}, {Rappaport}, {Podsiadlowski}, {Nelson},
  {Paxton}  \& {Todorov}}{{Lin} et~al.}{2011}]{lrp+11}
{Lin} J.,  {Rappaport} S.,  {Podsiadlowski} P.,  {Nelson} L.,  {Paxton} B.,
  {Todorov} P.,  2011, \mn@doi [\apj] {10.1088/0004-637X/732/2/70}, \href
  {http://adsabs.harvard.edu/abs/2011ApJ...732...70L} {732, 70}

\bibitem[\protect\citeauthoryear{{Lorimer} \& {Kramer}}{{Lorimer} \&
  {Kramer}}{2012}]{Handbook}
{Lorimer} D.~R.,  {Kramer} M.,  2012, {Handbook of Pulsar Astronomy}.
Cambridge University Press

\bibitem[\protect\citeauthoryear{{Lyne}}{{Lyne}}{1984}]{Lyne1984}
{Lyne} A.~G.,  1984, \mn@doi [\nat] {10.1038/310300a0}, \href
  {http://adsabs.harvard.edu/abs/1984Natur.310..300L} {310, 300}

\bibitem[\protect\citeauthoryear{{Manchester}, {Hobbs}, {Teoh}  \&
  {Hobbs}}{{Manchester} et~al.}{2005}]{Manchester05}
{Manchester} R.~N.,  {Hobbs} G.~B.,  {Teoh} A.,   {Hobbs} M.,  2005, VizieR
  Online Data Catalog, \href
  {http://adsabs.harvard.edu/abs/2005yCat.7245....0M} {7245}

\bibitem[\protect\citeauthoryear{{Mateo}, {Antoniadis}, {Tauris}  \& {van
  Kerkwijk}}{{Mateo} et~al.}{2017}]{Mateo17}
{Mateo} N.~M.,  {Antoniadis} J.,  {Tauris} T.~M.,   {van Kerkwijk} M.,  2017,
  in prep.

\bibitem[\protect\citeauthoryear{{Matthews} et~al.,}{{Matthews}
  et~al.}{2016}]{Matthews16}
{Matthews} A.~M.,  et~al., 2016, \mn@doi [\apj] {10.3847/0004-637X/818/1/92},
  \href {http://adsabs.harvard.edu/abs/2016ApJ...818...92M} {818, 92}

\bibitem[\protect\citeauthoryear{{Ng} et~al.,}{{Ng} et~al.}{2015}]{ncb+15}
{Ng} C.,  et~al., 2015, \mn@doi [\mnras] {10.1093/mnras/stv753}, \href
  {http://adsabs.harvard.edu/abs/2015MNRAS.450.2922N} {450, 2922}

\bibitem[\protect\citeauthoryear{{Ord}, {Bailes}  \& {van Straten}}{{Ord}
  et~al.}{2002}]{Ord02}
{Ord} S.~M.,  {Bailes} M.,   {van Straten} W.,  2002, \mn@doi [\apjl]
  {10.1086/342218}, \href {http://adsabs.harvard.edu/abs/2002ApJ...574L..75O}
  {574, L75}

\bibitem[\protect\citeauthoryear{{Phinney}}{{Phinney}}{1992}]{Phinney92}
{Phinney} E.~S.,  1992, \mn@doi [Philosophical Transactions of the Royal
  Society of London Series A] {10.1098/rsta.1992.0084}, \href
  {http://adsabs.harvard.edu/abs/1992RSPTA.341...39P} {341, 39}

\bibitem[\protect\citeauthoryear{{Phinney} \& {Kulkarni}}{{Phinney} \&
  {Kulkarni}}{1994}]{Phinney94}
{Phinney} E.~S.,  {Kulkarni} S.~R.,  1994, \mn@doi [\araa]
  {10.1146/annurev.aa.32.090194.003111}, \href
  {http://adsabs.harvard.edu/abs/1994ARA%26A..32..591P} {32, 591}

\bibitem[\protect\citeauthoryear{{Rappaport}, {Podsiadlowski}, {Joss}, {Di
  Stefano}  \& {Han}}{{Rappaport} et~al.}{1995}]{Rappaport95}
{Rappaport} S.,  {Podsiadlowski} P.,  {Joss} P.~C.,  {Di Stefano} R.,   {Han}
  Z.,  1995, \mn@doi [\mnras] {10.1093/mnras/273.3.731}, \href
  {http://adsabs.harvard.edu/abs/1995MNRAS.273..731R} {273, 731}

\bibitem[\protect\citeauthoryear{{Reid} et~al.,}{{Reid} et~al.}{2014}]{Reid14}
{Reid} M.~J.,  et~al., 2014, \mn@doi [\apj] {10.1088/0004-637X/783/2/130},
  \href {http://adsabs.harvard.edu/abs/2014ApJ...783..130R} {783, 130}

\bibitem[\protect\citeauthoryear{{Rickett}}{{Rickett}}{1977}]{Rickett77}
{Rickett} B.~J.,  1977, \mn@doi [\araa] {10.1146/annurev.aa.15.090177.002403},
  \href {http://adsabs.harvard.edu/abs/1977ARA%26A..15..479R} {15, 479}

\bibitem[\protect\citeauthoryear{{Savonije}}{{Savonije}}{1987}]{Savonije87}
{Savonije} G.~J.,  1987, \mn@doi [\nat] {10.1038/325416a0}, \href
  {http://adsabs.harvard.edu/abs/1987Natur.325..416S} {325, 416}

\bibitem[\protect\citeauthoryear{{Staveley-Smith} et~al.,}{{Staveley-Smith}
  et~al.}{1996}]{Staveley-Smith96}
{Staveley-Smith} L.,  et~al., 1996, \pasa, \href
  {http://adsabs.harvard.edu/abs/1996PASA...13..243S} {13, 243}

\bibitem[\protect\citeauthoryear{{Tauris}}{{Tauris}}{2011}]{Tauris11}
{Tauris} T.~M.,  2011, in {Schmidtobreick} L.,  {Schreiber} M.~R.,   {Tappert}
  C.,  eds,  Astronomical Society of the Pacific Conference Series Vol. 447,
  Evolution of Compact Binaries. p.~285 (\mn@eprint {arXiv} {1106.0897})

\bibitem[\protect\citeauthoryear{{Tauris} \& {Bailes}}{{Tauris} \&
  {Bailes}}{1996}]{tb96}
{Tauris} T.~M.,  {Bailes} M.,  1996, \aap, \href
  {http://adsabs.harvard.edu/abs/1996A%26A...315..432T} {315, 432}

\bibitem[\protect\citeauthoryear{{Tauris} \& {Savonije}}{{Tauris} \&
  {Savonije}}{1999}]{Tauris99}
{Tauris} T.~M.,  {Savonije} G.~J.,  1999, \aap, \href
  {http://adsabs.harvard.edu/abs/1999A%26A...350..928T} {350, 928}

\bibitem[\protect\citeauthoryear{{Tauris}, {van den Heuvel}  \&
  {Savonije}}{{Tauris} et~al.}{2000}]{tvs00}
{Tauris} T.~M.,  {van den Heuvel} E.~P.~J.,   {Savonije} G.~J.,  2000, \mn@doi
  [\apjl] {10.1086/312496}, \href
  {http://adsabs.harvard.edu/abs/2000ApJ...530L..93T} {530, L93}

\bibitem[\protect\citeauthoryear{{Tauris}, {Langer}  \& {Kramer}}{{Tauris}
  et~al.}{2011}]{tlk11}
{Tauris} T.~M.,  {Langer} N.,   {Kramer} M.,  2011, \mn@doi [\mnras]
  {10.1111/j.1365-2966.2011.19189.x}, \href
  {http://adsabs.harvard.edu/abs/2011MNRAS.416.2130T} {416, 2130}

\bibitem[\protect\citeauthoryear{{Tauris}, {Langer}  \& {Kramer}}{{Tauris}
  et~al.}{2012}]{Tauris12}
{Tauris} T.~M.,  {Langer} N.,   {Kramer} M.,  2012, \mn@doi [\mnras]
  {10.1111/j.1365-2966.2012.21446.x}, \href
  {http://adsabs.harvard.edu/abs/2012MNRAS.425.1601T} {425, 1601}

\bibitem[\protect\citeauthoryear{{Taylor}}{{Taylor}}{1992}]{Taylor92}
{Taylor} J.~H.,  1992, \mn@doi [Philosophical Transactions of the Royal Society
  of London Series A] {10.1098/rsta.1992.0088}, \href
  {http://adsabs.harvard.edu/abs/1992RSPTA.341..117T} {341, 117}

\bibitem[\protect\citeauthoryear{{Timmes}, {Woosley}  \& {Weaver}}{{Timmes}
  et~al.}{1996}]{tww96}
{Timmes} F.~X.,  {Woosley} S.~E.,   {Weaver} T.~A.,  1996, \mn@doi [\apj]
  {10.1086/176778}, \href {http://adsabs.harvard.edu/abs/1996ApJ...457..834T}
  {457, 834}

\bibitem[\protect\citeauthoryear{{Verbiest} et~al.,}{{Verbiest}
  et~al.}{2016}]{Verbiest2016}
{Verbiest} J.~P.~W.,  et~al., 2016, \mn@doi [\mnras] {10.1093/mnras/stw347},
  \href {http://adsabs.harvard.edu/abs/2016MNRAS.458.1267V} {458, 1267}

\bibitem[\protect\citeauthoryear{{Yao}, {Manchester}  \& {Wang}}{{Yao}
  et~al.}{2017}]{Yao17}
{Yao} J.~M.,  {Manchester} R.~N.,   {Wang} N.,  2017, \mn@doi [\apj]
  {10.3847/1538-4357/835/1/29}, \href
  {http://adsabs.harvard.edu/abs/2017ApJ...835...29Y} {835, 29}

\bibitem[\protect\citeauthoryear{{van Straten}}{{van
  Straten}}{2004}]{vanStraten04}
{van Straten} W.,  2004, \mn@doi [\apjs] {10.1086/383187}, \href
  {http://adsabs.harvard.edu/abs/2004ApJS..152..129V} {152, 129}

\bibitem[\protect\citeauthoryear{{van Straten}}{{van
  Straten}}{2006}]{vanStraten06}
{van Straten} W.,  2006, \mn@doi [\apj] {10.1086/501001}, \href
  {http://adsabs.harvard.edu/abs/2006ApJ...642.1004V} {642, 1004}

\bibitem[\protect\citeauthoryear{{van Straten}}{{van
  Straten}}{2013}]{vanStraten13}
{van Straten} W.,  2013, \mn@doi [\apjs] {10.1088/0067-0049/204/1/13}, \href
  {http://adsabs.harvard.edu/abs/2013ApJS..204...13V} {204, 13}

\bibitem[\protect\citeauthoryear{{van Straten}, {Manchester}, {Johnston}  \&
  {Reynolds}}{{van Straten} et~al.}{2010}]{vanStraten10}
{van Straten} W.,  {Manchester} R.~N.,  {Johnston} S.,   {Reynolds} J.~E.,
  2010, \mn@doi [\pasa] {10.1071/AS09084}, \href
  {http://adsabs.harvard.edu/abs/2010PASA...27..104V} {27, 104}

\bibitem[\protect\citeauthoryear{{van Straten}, {Demorest}  \& {Oslowski}}{{van
  Straten} et~al.}{2012}]{vanStraten12}
{van Straten} W.,  {Demorest} P.,   {Oslowski} S.,  2012, Astronomical Research
  and Technology, \href {http://adsabs.harvard.edu/abs/2012AR%26T....9..237V}
  {9, 237}

\makeatother
\end{thebibliography}


\bsp	
\label{lastpage}
\end{document}